# Supersonic Cluster Beam Fabrication of a Stretchable Keyboard with Multilayer Electrical Interconnects Integrated with Paper-based Flexible Devices

Andrea Bellacicca and Paolo Milani

CIMAINA and Dipartimento di Fisica, Università degli Studi di Milano,

via Celoria 16, 20133 Milano, Italy

**Abstract**

The development of stretchable and flexible electronics based on soft polymeric substrates requires novel approaches in the design, fabrication, integration, and packaging of passive and active electronic components. Here we demonstrate the fabrication, based on the use of supersonic cluster beam implantation and deposition, of a system consisting of a stretchable keyboard made by PDMS connected with a touch sensor and a matrix of LEDs printed on paper, through paper-based interconnections. The stretchable keyboard has conductive vias to electrically connect the top and bottom layers. Gold conductive films printed on paper are used to connect the keyboard with the electronic components that drive the LED matrix.

# 1. Introduction

The development of stretchable and flexible electronics based on soft polymeric substrates and components is pivotal for applications in wearable devices and sensors, personalized healthcare, soft robotics, and smart prosthetics [1, 2, 3]. The use of polymeric substrates requires a radical change in the design, fabrication, integration, and packaging of passive and active electronic components [3, 4, 5]. Novel approaches have, as a benchmark, well-established and economically sustainable production technologies developed in the last four decades for rigid substrates.

Complex stretchable electronic systems have been fabricated by embedding ultra-thin semiconductor-based integrated circuits in elastomeric matrices such as Polydimethylsiloxane (PDMS) [6]. This approach required the development of an electrical interconnectivity based on meander-like conductive paths resilient to compressive and tensile strain associated with stretching, bending and twisting [7]. Although very powerful and effective, this solution is highly complex and expensive since it requires the same facilities of silicon micro and nanofabrication.

An alternative to circuit embedding is the Direct Chip Attachment (DCA) where a microchip or die is directly mounted on, and electrically connected through a circuit printed on a flexible substrate (Chip on Flex, COF) [8, 9]. DCA typically deals with relatively large conductive features: low cost, speed of fabrication, and the ability to integrate multiple, discrete components into a functional system are the key parameters [3]. At present flexible substrate materials for COF are based on multilayer manufacturing and the attached discrete components are connected by wire bonding [9].

A recently proposed approach to produce stretchable electrical circuits is based on the use of polymer-metal nanocomposites produced by implanting neutral metallic clusters in PDMS by supersonic cluster beam implantation (SCBI) [10, 11]. Metal-polymer nanocomposites are very convenient as substrates for stretchable circuits, however their mechanical properties (Young modulus) are critical for the interconnection with other devices where the use of wires and rigid substrates is required.

Two major challenges are still associated with the fabrication of stretchable devices: i) the limitation to integration density of the components when only a single layer of interconnects is used; ii) the connection of stretchable parts with different hard components or substrates: the matching between soft and rigid components in electrical or mechanical connections remains a major technical issue to be solved.

A polymeric substrate very promising for the fabrication of flexible electronic circuits is paper [3, 12]: it is low cost and fully recyclable, it can be easily shaped and trimmed with scissors or cutters and it can be used for complex self-standing 3D structures, fluidics and electrochemical applications [13, 14]. Recently we have reported the fabrication of passive electrical components (resistors and capacitors) on plain paper by an additive and parallel technology consisting of supersonic cluster beam deposition (SCBD) coupled with shadow mask printing [15]. Compared to standard deposition technologies, SCBD allows for the rapid production of components with different shape and dimensions while controlling independently the electrical characteristics. Discrete electrical components produced by SCBD are very robust against deformation and bending and they can be easily assembled to build circuits with desired characteristics [15]. This suggests the use of paper for the fabrication of flexible electrical connection and substrates with suitable mechanical properties to be used in conjunction with soft stretchable components as well as with rigid substrates.

Here we demonstrate the fabrication, based on the use of SCBI/SCBD, of a system consisting of a stretchable keyboard made by PDMS connected with a touch sensor and a LED matrix printed on paper through paper-based interconnections. The stretchable keyboard has conductive vias to electrically link the front and bottom layers. Gold conductive films printed on paper are used to connect the keyboard with the electronic circuit that is in charge to drive the LED matrix.

## 2. System Fabrication and Characterization

### 2.1 Principle of operation and layout

A schematic diagram of the system consisting of a stretchable keyboard, a touch sensor, an Arduino board and a LED matrix is reported in Fig. 1. The stretchable keyboard is connected to a 12-key capacitive touch sensor MPR121 [16] through a flexible wiring made by printed electrically conductive gold paths on a paper substrate. An Arduino electronic board drives the capacitive sensors and runs the paper-based LED matrix according to the signals received from the touch sensor. The LED matrix is composed by nine LEDs powered by resistors printed on paper by SCBD and it is connected to the Arduino board by standard led-strip connectors.

The capacitance variation produced by touching a pad on the keyboard is transmitted to the rear side of the keyboard through conducting vias and then to the capacitive sensor connected by the metallized paper conductors. The sensor generates a set of univocal numbers identifying which pad has been touched and it sends them to the Arduino board through an Inter-Integrated Circuit ($I^2C$) communication channel. This triggers the Arduino board to turn on or off the correspondent LED fixed on the flexible paper substrate.

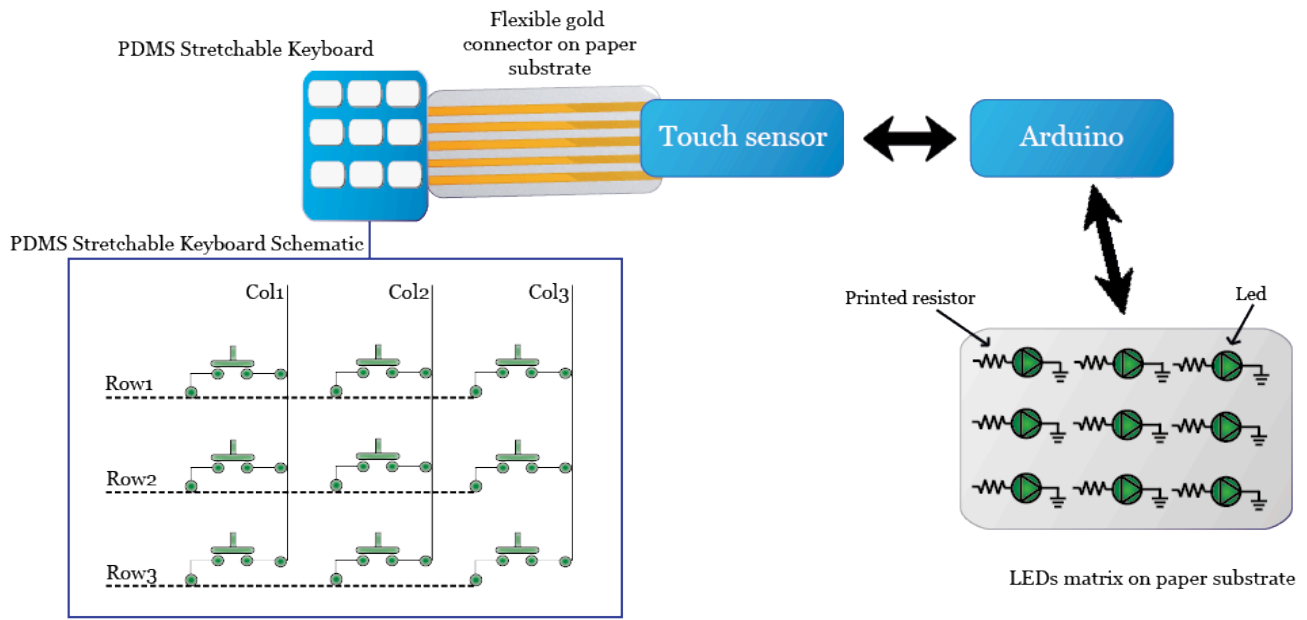

*Figure 1*: *Schematic representation of a PDMS stretchable keyboard connected to a touch sensor with flexible conductive electrodes realized on a paper substrate. The touch sensor sends data to an Arduino controller that turn on or off the LEDs on a matrix printed on a paper substrate. The inset shows the electrical scheme of the keyboard: columns and rows are printed on the opposite sides of the PDMS slab and connected by vias.*

The stretchable keyboard is based on a matrix architecture (Fig. 1) printed on both sides of the PDMS substrate. The bottom layer controls the columns of the matrix and it establishes the electrical connection between the keyboard and the paper connectors, the top layer integrates nine touchpads and electrical paths acting as bus dedicated to control the rows of the matrix. The two layers of the keyboard are electrically connected by twelve conductive vias. Every touchpad is realized splitting the pad in two separate parts: one connected to the column-control bus and the other connected to the row-control bus. The keyboard and the electrodes printed on paper are superimposed and bolted to a polylactic acid (PLA) plate to assure the electrical connection.

## 2.2 Printing of conductive paths and vias

Conductive paths (on PDMS and paper) and vias were fabricated by a Supersonic Cluster Beam Implantation/Deposition (SCBI/SCBD) apparatus equipped with a Pulsed Microplasma Cluster Source (PMCS) (Fig. 2). A detailed description of the SCBI/SCBD technology and of the deposition apparatus is reported in [10, 17]. Briefly neutral gold clusters are produced by the PMCS and accelerated for deposition or implantation in a supersonic expansion. A PMCS schematically consists of a ceramic body with a cavity where a target gold rod, acting as a cathode, is sputtered by a localized electrical discharge ignited during the pulsed injection of Ar gas at high pressure (40 bar). The sputtered metal atoms from the target thermalize with the carrier gas and aggregate in the source cavity forming metal clusters; the carrier gas-cluster mixture expands out of the PMCS through a nozzle into a low-pressure ($10^{-4}$ mbar) expansion chamber, thus producing a highly collimated supersonic beam with a divergence lower than 1°. The central part of the beam enters a second vacuum chamber (deposition chamber, at a pressure of about $10^{-5}$ mbar) through a skimmer and it impinges on the stencil mask-target substrate system. The cluster beam flux is monitored in real time by a quartz microbalance placed close to the substrate. A motorized samples holder allows large area deposition moving the substrate target in the perpendicular plane of the nanoparticles beam [10]. An additional servomotor rotates the substrate target along is vertical axis spanning 180 degrees with a resolution of ~9 degree.

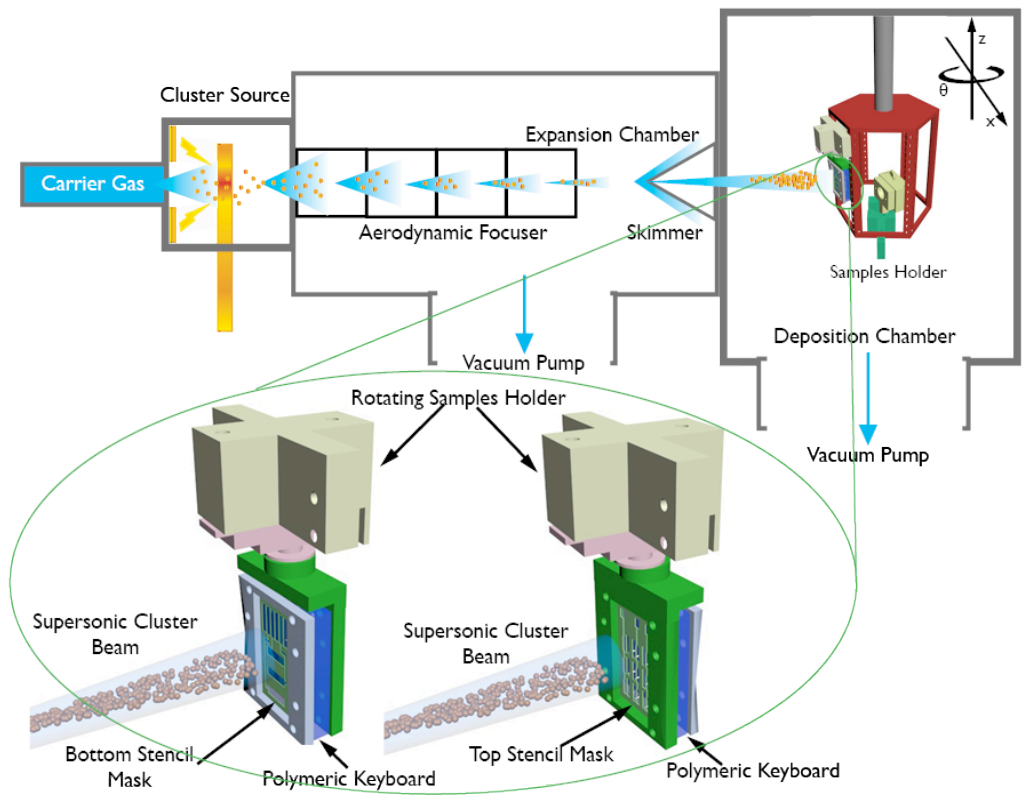

*Figure 2*: *Schematic representation (no to scale) of the SCBI/SCBD deposition apparatus. The inset shows the substrate holder used to fabricate on PDMS a double-sided electronic circuit electrically linked through metalized vias in a single-step process.*

In the case of soft PDMS substrate, we used Supersonic Cluster Beam Implantation (SCBI) [10, 11]. SCBI is based on the use of a highly collimated supersonic beam carrying metallic clusters with a kinetic energy of about 0.5 eV·atom-1. Even if the kinetic energy is significantly lower than in ion implantation, neutral clusters are able to penetrate up to tens of nanometers into the polymeric target forming a conducting nanocomposite and avoiding electrical charging and carbonization [11]. The amount of nanoparticles deposited onto the substrate target is measured in term of equivalent thickness defined as the thickness of a film made by an equivalent amount of nanoparticles deposited onto a rigid substrate [10]. We placed a partially masked Si wafer close to the substrate during the metallization process, after the deposition, we removed the mask and we measured the step of the cluster-assembled film using a P-6 stylus profilometer (KLA Tencor).

*2.3 Keyboard fabrication*

The keyboard (100 x 50 x 0.5 cm) has been printed on a PDMS substrate by implantation of gold clusters. PDMS was fabricated by mixing the monomer and the curing agent (Dow Corning, Sylgard 184) in a 10:1 ratio and stirred with a spatula for 20 minutes. Subsequently the mixture was put in a desiccator and degassed for one hour. The compound was then poured into the mold provided with twelve dowels pins to shape high smoothed holes to be used as conductive vias (Fig. 3A).

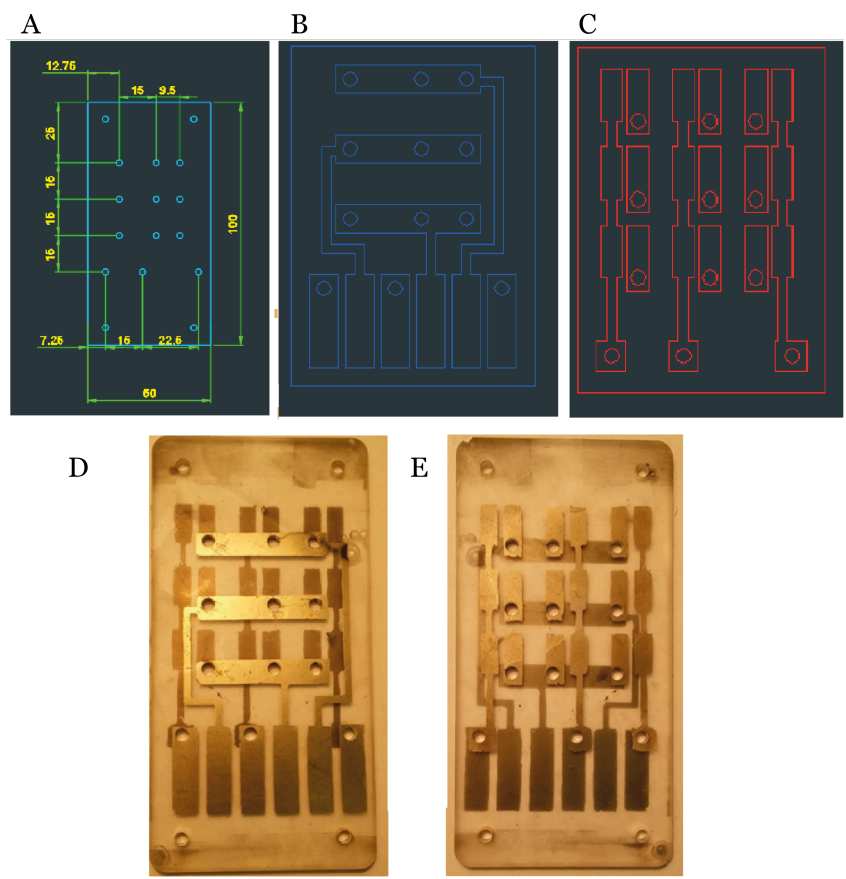

*Figure 3*: *A) Keyboard design: twelve holes are used as conductive vias to connect the electrodes on the top side of the keyboard with the electrodes on the bottom side. B) Top side stencil mask with*

*the nine touchpads used as buttons. C) Bottom side stencil mask with the six electrodes used to send the capacitive signal to the external electronic circuit. D) Picture of the keyboard bottom side after metallization. E) Picture of the keyboard top side, after metallization, with the conductive pads*

The PDMS substrate was sandwiched between two stencil masks reproducing the electrical paths to be printed and mounted on the rotating sample holder system (Fig. 2) in order to print the conductive electrodes on both sides of the target and the conductive vias during the same implantation process. The conductive paths were printed using SCBI combined with a rastering technique to ensure a homogeneous metallization [18]. First the top and bottom sides of the keyboard were deposited (Figs. 3B and 3C). The substrate was then rotated of roughly 30 degrees and 120 degrees respectively to metalize the twelve holes and to obtain the conductive vias that link the top and bottom sides of the keyboard. The angles were chosen to assure that the nanoparticles beam covers all the internal surface of the vias (see Fig. 4). The nine conductive pads and the bus controlling the columns were printed on the top side of the keyboard (Figs. 3C and 3E) while on the bottom side we printed the bus to control the rows and the links between the touchpads and the electrodes used to transfer electrical signals to the external control circuit (Figs. 3B and 3D).

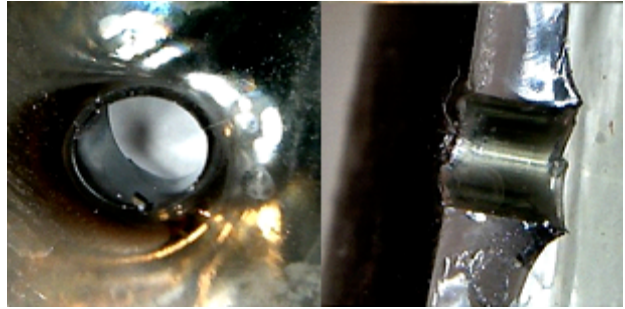

*Figure 4. Top side view and section of the PDMS vias metallized by SCBI.*

## 2.4 LED matrix fabrication

The LED matrix was realized using SCBD to deposit gold nanoparticles on a Xerox paper substrate (80 g/m$^2$) topped with a PLA stencil mask to shape the electrical paths [15]. The electrical resistance of the conductive paths was tailored by measuring in situ the amount of nanoparticles deposited onto the paper. High power LEDs were glued on the paper and connected to the paths through a conductive silver paste to establish a good electrical contact (Figs. 5A and 5B).

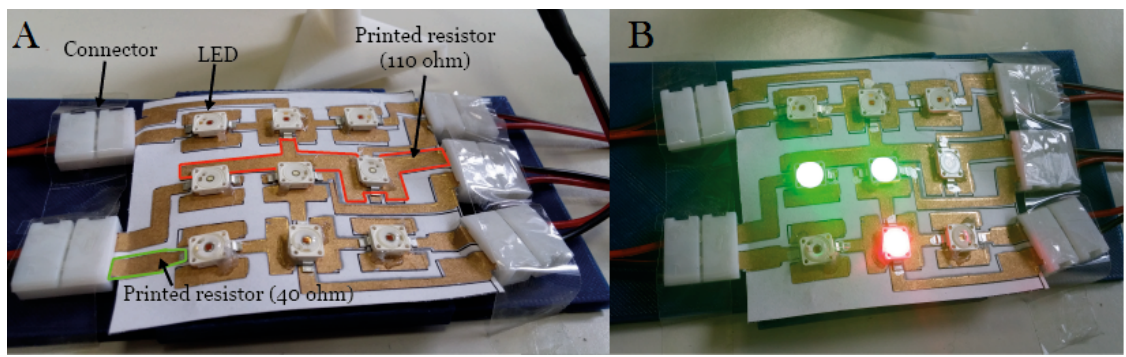

***Figure 5.*** *A) Picture of the LEDs matrix showing the electrical circuit fabricated by SCBD of gold clusters on paper and the glued LEDS. B) LEDs are turned on and off by touching the PDMS keyboard.*

*2.5 Gold/Paper conductive connectors*

      The connection between the electrodes of the bottom side of the PDMS keyboard and an external drive electronic circuit were made depositing gold with physical vapor deposition on a paper substrate topped with a PLA stencil mask to frame the electrical paths (Fig. 6).

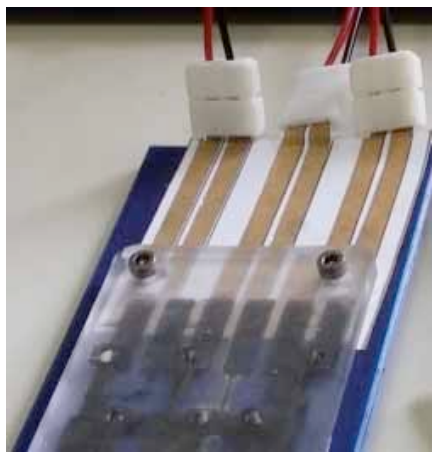

***Figure 6***: *Conductive paths on paper connecting the PDMS keyboards with standard connectors and wires.*

*2.6 Electro-mechanical characterization*

      The electrical resistance of the vias was measured using an Agilent 34410A under stretching cycles up 5% of linear deformation. In order to assess the durability of the electrical connection under repeated stress cycles, a portion of the keyboard of 10x20 mm including a vias were cut. The resistance was recorded after one hundred cycles of stretch/relax a 5% along the longest side.

# 3. Results and Discussion

The assembled device is shown in Fig. 7: touching the touchpads causes the correspondent LED to turning on or off. This test is replicated while bending, twisting and stretching the keyboard with no observation of delays or malfunctioning. The electrical connection between the keyboard and the electrodes printed on paper is robust enough to pipe the capacitive signal from the keyboard to the touch sensor, withstanding the mechanical deformation applied to the keyboard.

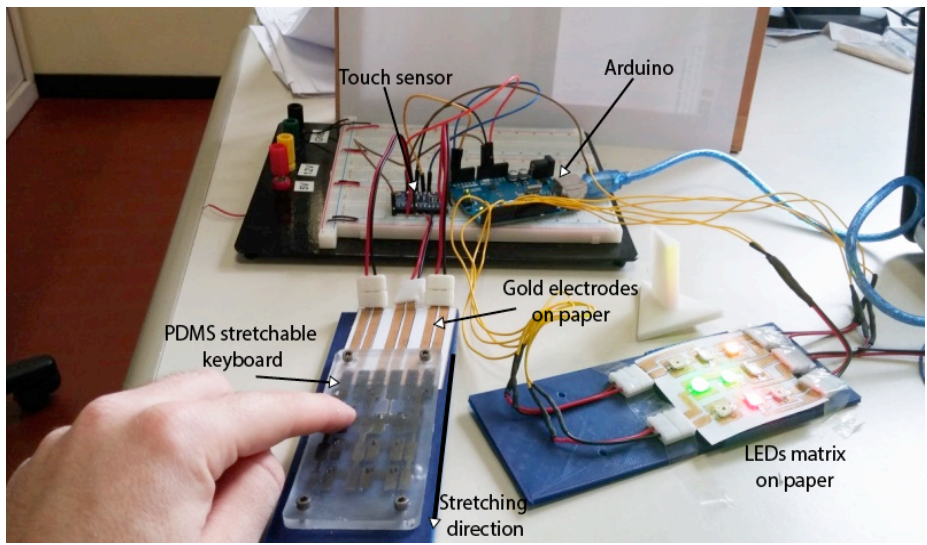

*Figure 7: Picture of the assembled device working under the finger pressure. The electric signal is red from the keyboard to the electronic circuit through a gold/paper flexible electrodes. The Arduino board in response to the touch turn on or off a LED in the LEDs matrix made of the flexible gold/paper circuit. The picture shows a test performed with the keyboard stretched of 5% along the direction indicated by the arrow.*

We used a matrix architecture for the keyboard in order to significantly simplify the electronic circuit needed to control it. In fact, a direct connections scheme (one pin per one pad) would require a one-to-one link between the touchpads and the capacitive sensor inputs. Cleary, this is not feasible mainly because of two reasons: i) this scheme would require a touch sensor with the same number of input as the number of touchpads and the higher the amount of pins, the higher the complexity of the internal logic and thus the higher the price of the electronic circuit; ii) this electrical arrangement would have required a high numbers of conductive paths, leading to a considerable increasing in the final complexity of the whole electronic circuit.

On the contrary, a matrix architecture significantly reduces the design complexity allowing to control up to $N^2$ touchpads with just 2N sensor pins at the simple cost of using a two-layers electronic circuit. One layer has a bus dedicated to control the rows of the matrix; the second handles the columns of the matrix. This implies that it is possible to deal with up 36 independent touchpads with a 12-key touch sensor.

The matrix architecture is implemented in a device using a double-sided electronic circuit with conductive through-hole vias links. While this is of course trivial for standard electronic circuits made on hard substrates, on the other hand the realization of conductive vias in elastomeric substrates is still challenging with common metallization techniques [4, 19].

By using SCBI it is possible to fabricate conductive vias in PDMS thick layers, the direct metallization of the inner wall of the vias is obtained by exploiting the high collimation of the cluster to be implanted in the beam typical of supersonic expansions [20]. By tilting the substrate against the cluster beam it is possible to uniformly implant the clusters. The formation of a polymer-metal nanocomposite assures the resilience of the conducting film against stretching and bending [11].

As showed in Fig. 8 the electrodes and vias are still conductive during the stretching phase undergoing a small increase in the resistance values and restoring the initial value when the keyboard is reset to rest position.

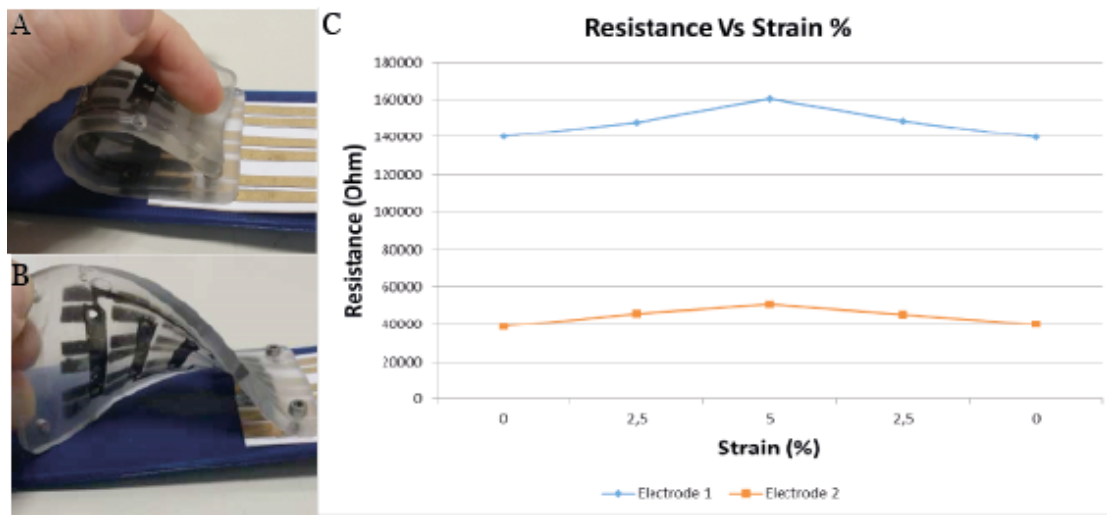

*Figure 8 A-B) stretching and bending of the PDMS keyboard. C) Evolution of the resistance for two of the conductive path-vias-conductive path of the keyboard during a stretching cycle.*

In order to realize the LED matrix on paper, we exploited the capability of directly printing on paper resistors by SCBD [15]. Standard electronic circuits are made by discrete components like resistors, capacitors, transistors, etc., or integrated components like microprocessors, electrically soldered on rigid printed circuit boards (PCB). The components are connected through conductive paths (CPs) usually made of copper. This architecture implies a net separation between components and electrical paths. For example, a typical circuit to power a LED is assembled using the following scheme: CP-resistor-CP-LED-CP.

Exploiting SCBD resistors and capacitors with well-defined electrical characteristic can be deposited on paper simply by controlling the amount of deposited nanoparticles [15]. This allows to fabricate conductive paths with controlled resistance value (resistive conductive path or RCP) to limit the amount of current that flows in the LEDs and demonstrating an alternative circuit scheme such as: RCP-LED-RCP. This provides a substantial simplification in the design and realization of electronic circuits on paper.

In the case of the LEDs matrix, the resistance of the gold elements printed of paper spanned from 100 ohm to 250 ohm obtained with a cluster-assembled gold film with an equivalent thickness of 180nm (Fig. 7). These resistance values guarantee that the current sunk from the Arduino board doesn't exceed the maximum value, attested around 200 mA.

## 4. Conclusions

In summary we reported the fabrication of a hybrid device based on the integration of electronic components and electrical circuits on different stretchable and flexible polymeric materials such a PDMS and paper. By using supersonic cluster beam implantation and deposition it is possible to fabricate conductive paths and vias on polymeric stretchable substrates able to sustain deformations without altering their electrical properties. The use of electrically conductive paths printed on paper allowed the connection of PDMS devices with classical "hard" electronic units and the fabrication of complex devices matrices. This work demonstrates that SCBI/SCBD is an enabling technology for the fabrication of a novel class of complex devices based on non-conventional substrates.